\documentclass[12pt]{iopart}
\usepackage{graphicx}
\begin{document}
\title{Deformation of the $CP(1)$ model leading to fixed size solitons in 2+1 dimensions}
\author{A J Peterson}
\address{School of Physics and Astronomy, University of Minnesota, Minneapolis, MN 55455, USA}
\ead{pete5997@umn.edu}

\begin{abstract}
We discuss static particle-like solitons in the 2+1 dimensional $CP(1)$ model with a small mass deformation $m$ preserving a $U(1) \times Z_2$ symmetry in the Lagrangian.  Due to the breaking of scale invariance, the energy function becomes a strictly increasing function of the soliton size $\rho$, and therefore no classical finite size solution exists in this model.  To remedy this we employ a well known technique of introducing a forth-order derivative term in the Lagrangian to force the soliton action to diverge at small values of $\rho$.  With this additional term the action exhibits a stable minimum at fixed size $\rho$.
\end{abstract}
\submitto{\JPA}
\maketitle

\section{Introduction}

The $CP(N-1)$ model has proven to be an indispensable tool for analyzing many interesting phenomena in high energy physics where it has consistently appeared as a laboratory for studying non-perturbative effects in four-dimensional Yang-Mills theories \cite{Witten:1978bc}\cite{Novikov:1984ac}.  The model in two dimensions shares many characteristics of four-dimensional Yang-Mills theories including asymptotic freedom \cite{Polyakov:1975rr} and infrared strong coupling.  The lower dimensionality of the theory provides a less hostile environment for studying the non-perturbative effects that appear in Yang-Mills theories where a strong coupling analysis is difficult to perform.  In addition to the qualitative analogies with Yang-Mills theories, the $CP(N-1)$ models and their variants tend to appear on the low energy dynamics of the gapless excitations of topological defects due to the spontaneous breaking of non-Abelian symmetries (see for example \cite{Shifman:2004dr} and \cite{Nitta:2013mj}).

The two-dimensional $CP(1)$ model in particular allows for the study of instantons \cite{Belavin:1975fg} and particle-like solitons in the extension to $2+1$ dimensions \cite{Woo:1976hu}.  A particular triumph of the instanton calculus worth mentioning is the exact calculation of fermion two point functions in the two-dimensional supersymmetric $CP(1)$ model \cite{Novikov:1984ac}.  This is in close relation to the calculation of the gluino vacuum condensate in supersymmetric gluodynamics \cite{Novikov:1983ee}.  New difficulties arise however when matter is included due to the loss of scale invariance.  The introduction of a Higgs sector in Yang-Mills theories results in the lifting of moduli from the instanton measure, which results in the suppression of instantons with sizes larger than the inverse Higgs mass scale \cite{'tHooft:1976fv}\cite{Novikov:1985ic}.  Such a suppression is also expected to occur in the two-dimensional $CP(1)$ model with a mass deformation, however to our knowledge no such calculation has been performed.  

Static particle-like solitons in the $2+1$-dimensional $CP(1)$ mass deformed models occur with similar complications since their sizes are not stabilized at finite values even when the mass term is absent (although see \cite{Leese:1991hr}).  This presents difficulties for quantization unless the soliton size can be fixed at finite value.  We wish to address this problem for the specific case of the non-supersymmetric $CP(1)$ model in $2+1$ dimensions with a small mass deformation preserving a $U(1) \times Z_2$ symmetry.  This choice of mass deformation is relevant to the $CP(N-1)$ model emerging as a moduli theory on the world sheet of flux tubes presenting string like solitons in four-dimensional Yang-Mills theories with a color-flavor locking mechanism $SU(N)_C \times SU(N)_F \rightarrow SU(N)_{diag}$ \cite{Shifman:2004dr}\cite{Gorsky:2004ad}-\cite{Hanany:2004ea}.  In this case the mass deformation is introduced at the four-dimensional level as an adjustable parameter, which preserves a $U(1)^{N-1} \times Z_N$ symmetry on the moduli space. 

The static solitons in the $2+1$-dimensional $CP(1)$ model can of course be lifted from the instantons in the two-dimensional case.  As we have mentioned above, we will consider the former as our physical context.   The reasoning for this choice is related to the method of fixing the size.  In the two-dimensional case the standard approach is to use the technique of constrained instantons first introduced in \cite{Frishman:1978xs} and \cite{Affleck:1980mp}, whereby a constraint on the functional measure is introduced forcing the size to assume a finite value, which is subsequently integrated over in the calculation of observable quantities.  A well known method of constrained instantons is through the valley method \cite{Balitsky:1986qn}\cite{Balitsky:1985in}, however there is no unambiguous choice of constraint criteria, and one may choose the constraint at will based on the relevance to the observable being calculated \cite{Vainshtein:1981wh}.  For the $CP(1)$ model, however, such an approach has the added difficulty that the mass term in the action diverges logarithmically when the original instanton solution is inserted in the action, and it is unclear how to regulate this term unambiguously.  We thus proceed to discuss a technique that is suited to the case of static particle-like solitions in $2+1$-dimensions.  

To approach this we follow the procedure illustrated in \cite{Skyrme:1961vq}-\cite{Faddeev:1996zj} where it was shown that in the four-dimensional Skyrme-Faddeev model a higher order derivative term was required to ensure the existence of finite size solitons in three dimensions under a scale transformation.  In this case the added term presents the Hopf invariant, which determines the topological degree of mapping $S_3 \rightarrow S_2$.  The topological non-trivial nature of the soliton solutions characterized by the Hopf invariant establishes their stability \cite{Faddeev:1996zj}\cite{Bolognesi:2007ut}.  

We introduce a similar higher order derivative term to the $2+1$-dimensional mass deformed $CP(1)$ model.  We observe that the added term causes the energy to diverge at small values of the size and thus the size becomes fixed at a finite value as in \cite{Faddeev:1996zj}.  This method was first applied to solitons in the $2+1$-dimensional $CP(1)$ model in \cite{Leese:1989gi} where the authors considered an alternative form of the deformation potential.

Throughout the analysis we will be working in the context of static solitons in $2+1$ dimensions, emphasizing that technically at the classical level they are the instantons in the two-dimensional case.  We will begin with a brief review of the instantons in the undeformed $O(3)$ model (sect. 2).  The small mass deformation will then be introduced and the consequences for the soliton size will be determined in section 3.  In section 4 the model will be modified to include a forth-order derivative term that supports solitons with fixed size.  In particular, we will show that for appropriately chosen coefficients the soliton can be constrained at finite size much less than the inverse mass.  We will conclude with a qualitative interpretation of our results and compare with similar calculations performed in four-dimensional Yang Mills theories.  For the present purposes we will remain strictly in the classical regime, leaving the quantization of the model for future discussions.

\section{Instantons in the $O(3)$ Sigma Model}

We will begin by reviewing the instanton solutions in the 2-Dimensional $O(3)$ sigma model, which present the static solitons in $2+1$ dimensions.  This brief analysis will follow that given in \cite{Shifman:2012zz} where more details can be found.

The Euclidean action for the original $O(3)$ sigma model can be written in the geometric representation as
\begin{equation}
S = \frac{2}{g^2}\int d^{2}x \frac{\partial_{\mu}\overline{\phi}\partial_{\mu}\phi}{(1+\overline{\phi}\phi)^2}.
\label{cp1action}
\end{equation}
Performing the Bogomol'nyi completion on this action we find
\begin{eqnarray}
\fl S = \frac{1}{g^2}\int d^{2}x \left[ \left(\partial_{\mu}\overline{\phi} \mp i \varepsilon_{\mu\nu}\partial_{\nu}\overline{\phi}\right)\left(\partial_{\mu}\phi \pm i \varepsilon_{\mu\rho}\partial_{\rho}\phi\right)\right. \nonumber \\
\left. \mp 2i\varepsilon_{\mu\nu}\partial_{\mu}\overline{\phi}\partial_{\nu}\phi\right](1+\overline{\phi}\phi)^{-2}.
\label{bogomolnyi}
\end{eqnarray}
It can be shown that the term in the second line of (\ref{bogomolnyi}) is a total derivative and thus represents the topological term in the action,
\begin{equation}
S = \frac{1}{g^2}\int d^{2}x\left(\partial_{\mu}\overline{\phi} \mp i \varepsilon_{\mu\nu}\partial_{\nu}\overline{\phi}\right)\left(\partial_{\mu}\phi \pm i \varepsilon_{\mu\rho}\partial_{\rho}\phi\right) + \frac{4\pi n}{g^2},
\label{Qaction}
\end{equation}
where $n$ is the topological charge:
\begin{equation}
n = \frac{1}{\pi}\int d^2x \left\{ \frac{d\overline{\phi}}{d\overline{z}}\frac{d\phi}{dz}-\frac{d\overline{\phi}}{dz}\frac{d\phi}{d\overline{z}} \right\}(1+\overline{\phi}\phi)^{-2}, \; z = x_1 + i x_2.
\label{ncharge}
\end{equation}
Choosing the upper sign in (\ref{Qaction}) we can see that the action achieves a local minimum if
\begin{equation}
\partial_{\mu}\phi+i\varepsilon_{\mu\nu}\partial_{\nu}\phi = 0,
\label{eqofmo}
\end{equation}
which is simply the Cauchy-Riemann condition 
\begin{equation}
\frac{d\phi}{d\overline{z}} = 0.
\label{cauchyriemmann}
\end{equation}
Thus solutions of (\ref{cp1action}) are in fact analytic functions of $z$.  In addition to ensure the action is finite the solutions $\phi(z)$ must be meromorphic functions.  In addition in the limit $|z|\rightarrow\infty$, $\phi(z)$ must be independent of the angular direction.  This compactifies the two dimensional Euclidean space to the sphere $S^2$.  Thus the solutions $\phi(z)$ represent topologically non-trivial mappings of the compactified Euclidean space $S^2$ to the target space $S^2$.  The number of poles in the solution $\phi(z)$ determines the topological charge.  The instanton with topological charge $n = 1$ can be represented as
\begin{equation} 
\phi(z) = \frac{\rho}{z-z_0},
\label{instanton}
\end{equation}
with action $S_0 = 4\pi/g^2$.  Here $\rho$ is a complex parameter with $|\rho|$ representing the instanton ``size", while $z_0$ determines the instanton center.  The scale parameter $\rho$ is arbitrary in the instanton action.  This is because scale invariance is unbroken at the classical level \cite{Belavin:1975fg}\cite{Shifman:2012zz}.

We mention for completeness that once the classical instanton is determined it is possible to show that the instanton measure with quantum corrections at one-loop order takes the form \cite{Jevicki:1977yd}\cite{Din:1979zv}
\begin{equation}
d\mu_{inst}=const \times M_{uv}^2 d^2 z_0\frac{d\rho}{\rho} \exp{(-S_0)},
\label{measure}
\end{equation}
where $M_{uv}$ is the ultraviolet cutoff of the model.  The instanton measure diverges logarithmically in the large $\rho$ limit due to the infrared strong coupling of the model.  It is expected that the introduction of a mass deformation $m$ will suppress instantons with sizes $\rho \sim 1/m$.  Corrections to this measure can be calculated using the procedure of constraint instantons \cite{Frishman:1978xs}\cite{Affleck:1980mp}, although we point out possible complications arising due to the divergent nature of the mass term (see (\ref{mcp1action}) below) when calculated using the instanton solution (\ref{instanton}).

At this point we abandon further discussion in the two-dimensional context and will now consider the instanton solutions as embedded in the $2+1$-dimensional case presenting the static solitons of interest.  We will consider only the time independent case and therefore the action $S$ in two dimensions corresponds to the energy $E$ in $2+1$ dimesions:
\begin{equation}
S(\rho) \rightarrow E(\rho)
\label{Conversion}
\end{equation}
In addition the coupling $2/g_0^2$ will acquire a mass dimension and we will relabel it:
\begin{equation}
\frac{2}{g_0^2} \rightarrow \xi.
\end{equation}

\section{Solitons in the $2+1$-Dimensional $O(3)$ Model with a Mass Deformation}

Our modification of the $O(3)$ sigma model is to introduce a mass term in the numerator of (\ref{cp1action}):
\begin{equation}
E = \xi\int d^{2}x \frac{\partial_t\overline{\phi}\partial_t\phi+\partial_i\overline{\phi}\partial_i\phi+|m|^2\overline{\phi}\phi}{(1+\overline{\phi}\phi)^2},
\label{mcp1action}
\end{equation}
where for our purposes we will assume $m$ is real.  Mass terms of this form appear in supersymmetric sigma models with twisted mass \cite{AlvarezGaume:1983ab}\cite{Gates:1984nk}.

The difficulty with the added term in (\ref{mcp1action}) is that a simple substitution of the original solution (\ref{instanton}) leads to a divergent integral.  This difficulty can be surmounted if we lift the multi-instanton solution from two-dimensions and attach a stationary time dependence \cite{Leese:1991hr}:
\begin{equation}
\phi(z,t) = \sum_{k=1}^n \frac{\rho_k}{z-z_k} e^{imt},
\label{stationary}
\end{equation}
where $\rho_k$ are complex numbers subject to the polygonal constraint 
\begin{equation}
\sum_k \rho_k = 0.
\label{cconstraint}
\end{equation}
It can be shown that (\ref{stationary}) satisfies the equations of motion.  The solution is characterized by the topological charge $n$ as well as a Noether charge $Q$ from the $U(1)$ symmetry:
\begin{equation}
Q  =\xi\int d^2x i \frac{\phi\partial_t\overline{\phi}-\overline{\phi}\partial_t{\phi}}{(1+\overline\phi\phi)^2}.
\label{Qcharge}
\end{equation}
The conservation of $Q$ from the $U(1)$ symmetry protects the soliton size from rolling to zero.  In the limiting case of the $n$-instanton configuration 
\begin{equation}
\phi(z,t)=\left(\frac{\rho}{z}\right)^ne^{imt},
\label{ninstanton}
\end{equation}
we can immediately determine the Neother charge:
\begin{equation}
Q = \xi \frac{2 \pi^2m}{n^2}|\rho|^2 \left(\sin{\frac{\pi}{n}}\right)^{-1}
\end{equation}
Using (\ref{ncharge}), (\ref{mcp1action}), (\ref{ninstanton}), and (\ref{Qcharge}) the minimal energy is
\begin{equation}
E = 2\pi\xi |n| + m|Q|.
\label{stationaryE}
\end{equation}
This solution is a stationary soliton known as a Q-lump which was first introduced in \cite{Leese:1991hr} where more details regarding quantization, stability, and interactions can be found.  We note however that the Noether charge $Q$ diverges in the infrared limit for $n=1$ and thus the stationary solitons only exist for $n > 1$.  This divergence is logarithmic due to the power like behavior of (\ref{ninstanton}).  In what follows we would like to find a static solution for a single topological winding that avoids this divergent behavior.

Returning to the static case, we expect the solution (\ref{instanton}) to be valid in the limit of $|z| << 1/m$.  However, for values of $|z|$ beyond $1/m$ we expect the soliton is modified to show a cutoff at $1/m$.  It is therefore assumed that the soliton will take on a form similar to
\begin{equation}
\phi(z) \approx \frac{\rho}{z-z_0}\exp{(-m|z-z_0|)} \times P(m|z-z_0|),
\label{minstanton}
\end{equation} 
where $P(m|z-z_0|) \rightarrow 1$ as $|z-z_0| \rightarrow 0$ is a polynomial function of its argument.  To simplify notation we will define $r \equiv |z-z_0|$ and $u \equiv m|z-z_0|$.  In addition we will use a dimensionless parameter $\lambda$ defined as
\begin{equation}
\lambda \equiv m\rho.
\label{lambdadefinition}
\end{equation}

The exponential in (\ref{minstanton}) acts as an infrared cutoff in (\ref{mcp1action}) at $r \sim 1/m$.  In the logarithmic approximation we may use the form of the solution (\ref{instanton}) and impose a limit of integration $r \rightarrow 1/m$.  This results in an approximate form of the energy:
\begin{equation}
E \rightarrow 2\pi\xi\left(1+\lambda^2\log{\frac{1}{\lambda}}\right).
\label{sapprox}
\end{equation}
Comparison of this approximation with a trial function minimization of the action for several values of $\lambda << 1$ is shown in Figure 1.  Not surprisingly, this solution exhibits a infrared divergent logarithm in the second term similar to the divergence of the energy in (\ref{Qcharge}) and (\ref{mcp1action}) for the $n=1$ Q-lump.  In our case however, the divergent logarithm is naturally regulated with a $\rho$ dependence and thus the complete term is finite.  This is of course due to the exponential cutoff of (\ref{minstanton}).
\begin{figure}[h!]
\centering
\includegraphics{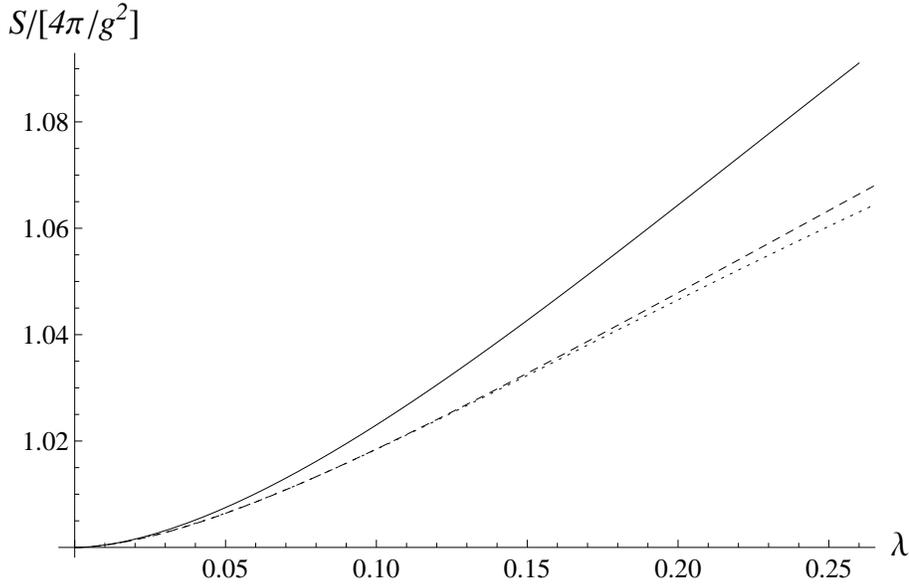}
\caption{Here the solid line is the approximate energy (\ref{sapprox}) as a function of $\lambda$.  The dashed line is the energy plotted as a function of $\lambda$ by minimization using the solution (\ref{minstanton}) with the trial function $P(u) =1+ a(\lambda) u$.  The dotted line shows a similar minimization of the energy with trial function $P(u) =1+ a(\lambda) u+b(\lambda)u^2$.  Marginal improvement is observed with the additional quadratic term in $P(u)$ for the range of $\lambda$ considered.}
\end{figure}

Clearly the soliton (\ref{minstanton}) breaks the invariance of the energy to the size $\rho$.  In addition this causes the minimal energy to tend towards solutions with $\rho \rightarrow 0$.  Thus the stationary soliton for finite size $\rho \neq 0$ does not exist in the classical limit.

\section{Supporting the Stationary Soliton at Finite Size}

To ensure the existence of solutions with finite size we are forced to include new terms that will result in a minimal energy at finite size $\rho$.  A typical modification to the O(3) model when considering stationary solitons is to include a term that is forth order in the derivative \cite{Skyrme:1961vq}-\cite{Faddeev:1996zj}\cite{Manton:2004tk}.  Written in the form of the field $\vec{S}(x)$ the term takes on a form \cite{Bolognesi:2007ut}:
\begin{equation}
\delta E = \xi\int d^2x\frac{\beta}{8}\left(\partial_{\mu}\vec{S}\times\partial_{\nu}\vec{S}\right)\cdot\left(\partial_{\mu}\vec{S}\times\partial_{\nu}\vec{S}\right),
 \label{higherdlagrangian}
\end{equation} 
where the parameter $\beta << \rho^2 << 1/m^2$ is assumed small enough to neglect corrections from the mass $m$ in (\ref{higherdlagrangian}).  Thus for this portion of the energy we will assume that (\ref{eqofmo}) is valid.  In this approximation (\ref{higherdlagrangian}) reduces to
\begin{equation}
\delta E \approx \xi \int d^2x \beta \frac{(\partial_{\mu}\overline{\phi}\partial_{\mu}\phi)^2}{(1+\overline{\phi}\phi)^4},
\label{higherdreduced}
\end{equation}

Including the additional forth order derivative term in the energy (\ref{mcp1action}) and inserting the soliton solution (\ref{instanton}), where we are ignoring the mass correction at large $|z|$ for this term, we arrive at the following approximation to the energy:
\begin{equation}
E(\rho, \beta) \approx 2\pi\xi\left(1+m^2\rho^2\log{\frac{1}{m\rho}} + \frac{2}{3}\frac{\beta}{\rho^2}\right)
\label{modifiedaction}
\end{equation} 
To simplify notation we define another dimensionless parameter 
\begin{equation}
\alpha \equiv \frac{2}{3}\beta m^2.
\label{alphadefinition}
\end{equation}  
Thus we find,
\begin{equation}
E(\lambda, \alpha) \approx 2\pi\xi\left(1+\lambda^2\log{\frac{1}{\lambda}}+\frac{\alpha}{\lambda^2}\right).
\label{lambdaalphaaction}
\end{equation}

It is now clear that the energy (\ref{lambdaalphaaction}) admits a minimum at a finite value of $\lambda$ defined parametrically by: 
\begin{equation}
\lambda_{0}^4\left(\log{\frac{1}{\lambda_{0}}}-\frac{1}{2}\right) = \alpha.
\label{lambdamin}
\end{equation} 
Thus the soliton is stabilized at a small but finite value of $\lambda = m\rho$ as can be seen in Figure 2.  In addition a plot of the minimized energy as a function of $\beta$ is included in Figure 3.  We see that the approximate energy (\ref{lambdaalphaaction}) agrees well with the energy minimized with the approximate solution (\ref{minstanton}).  The value of $\lambda_0$ determined by (\ref{lambdamin}) implies that the energy is minimized at $\rho_0^2 \sim \sqrt{\beta}/m$, and thus the requirement that $\rho_0^2 \ll 1/m^2$ is satisfied.
\begin{figure}[h!]
\centering
\includegraphics{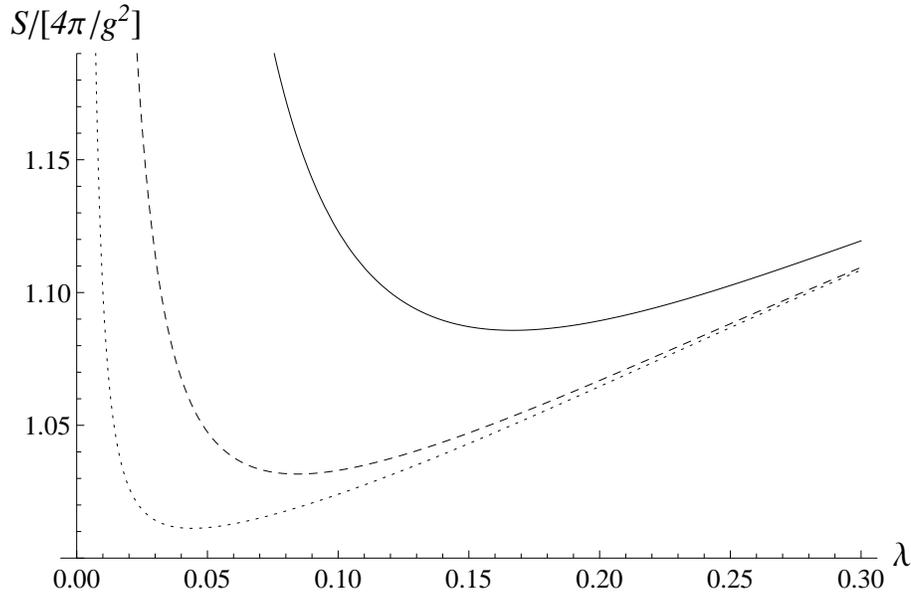}
\caption{The energy $E$ is plotted as a function of $\lambda$ various values of $\alpha$.  The solid, dashed, and dotted lines are plotted for values $\alpha$ at 0.001, 0.0001, and 0.00001 respectively.}
\end{figure}
\begin{figure}[h!]
\centering
\includegraphics{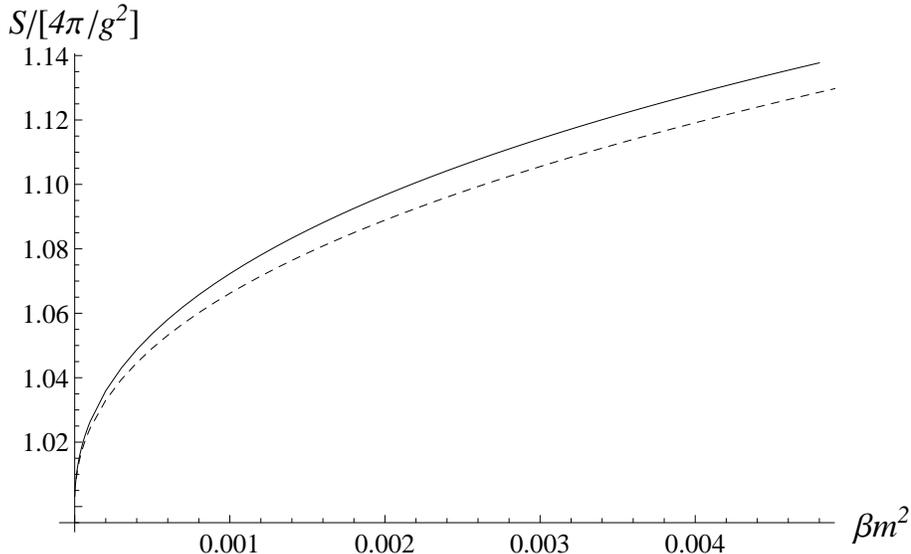}
\caption{Here we have plotted the minimal value of the energy as a function of $\beta m^2$.  Again the solid line shows the result using energy in (\ref{lambdaalphaaction}), while the dashed line shows the result using the approximate solution (\ref{minstanton}) with the trial function $P(u) =1+ a(\lambda)u$.}
\end{figure}

It is straightforward to continue this analysis to higher topological windings $n > 1$.  In this case we find for the energy:
\begin{equation}
E \approx 2\pi\xi\left\{|n|+\frac{\pi}{2n^2}\lambda^2\left(\sin{\frac{\pi}{n}}\right)^{-1} +\alpha\frac{\pi(n^2-1)}{2}\frac{1}{\lambda^2}\left(\sin{\frac{\pi}{n}}\right)^{-1}\right\},
\label{nsolitons}
\end{equation}
with a minimal energy at
\begin{equation}
\lambda_{(0,n)}^2 \approx \sqrt{\alpha n^2(n^2-1)}.
\label{nlambdamin}
\end{equation}
The requirement that $\rho_0^2 \ll 1/m^2$ is preserved for small $n$.  We point out that the second term in (\ref{nsolitons}) for small $\lambda$ is precisely half of the corresponding term in (\ref{startionaryE}) for the stationary case.

\section{Discussion and Conclusions}
We have shown that the $CP(1)$ model with a $U(1) \times Z_2$ preserving mass deformation admits the existence of static solitons with unit winding.  This conclusion is not surprising since the target space of the mass deformed $CP(1)$ model has the same topology $S_2$ as the original $O(3)$ sigma model.  Therefore we should still expect solutions of the equations of motion that provide a nontrivial mapping from the $S_2$ Euclidean space-time onto the target space in the mass deformed case.  We observe that the soliton size will roll to zero under time evolution due to the loss of scale symmetry.  In order to fix the size $\rho$ at a finite value we've included a higher order derivative term in the Lagrangian similar to the higher derivative term introduced in the Skyrme-Faddeev model \cite{Faddeev:1996zj}.  Fixing the soliton size allows for quantization to take place.

We have mentioned that solitons in the $2+1$-dimensional $CP(1)$ model have already been discovered and analyzed in \cite{Leese:1991hr} where the soliton solutions are given a stationary $U(1)$ time dependence and topological winding $n > 1$.  Their sizes are fixed (but remain arbitrary) by the conservation of the Noether charge $Q$.  In that case the power law dependence of the soliton solution prevents finite energy solution with unit winding to exist due to the infrared divergence.  The static soliton case we've considered allows for single winding solutions to exist.  We admit these solutions are not as elegant as the stationary solutions of \cite{Leese:1991hr} because the size fixing is not protected by the $U(1)$ symmetry in the static case.  Instead our model imposes the size fixing as a minimization of the energy function.

As emphasized above we've chosen to discuss soltions of the mass deformed $CP(1)$ model in the context of $2+1$-dimensions.  The alternative context of instantons in the two-dimensional $CP(1)$ model with a mass deformation would be relevant for comparison to the case of instantons in four-dimensional Higgsed Yang-Mills theories.  In the Yang Mills theory the instanton solution for the gauge field $A_{\mu}^a(x)$ is assumed to not change when the scalar field $\chi^{i}(x)$ is introduced.  This is justified at weak coupling.  The field $\chi^{i}(x)$ is determined by minimization of the action with the instanton solution for $A_{\mu}^a(x)$.  The source term induced by the nonzero scalar field can be made negligible by considering constraint instantons \cite{Frishman:1978xs}\cite{Affleck:1980mp} with sizes that satisfy this approximation.  As seen above, in the massive $CP(1)$ model this approach runs into difficulty because the original instanton solution (\ref{instanton}) causes the additional term in (\ref{mcp1action}) to diverge in the infrared.  Aside from a few general remarks we have made no attempt to analyze this context in detail.

We would like to conclude by noting some suggestions for future research.  We have confined our discussion of solitons in the mass deformed $CP(1)$ model in $2+1$ dimensions to the classical level.  Some discussion outlining the requirements for soliton quantization have been made, however we have not carried out a detailed calculation.  In addition we have avoided discussing soliton interactions and scattering processes for the static case.  We have made a few qualitative remarks regarding the corresponding instantons in the two-dimensional model, however we leave a detailed exploration of the constraint method of \cite{Frishman:1978xs} and \cite{Affleck:1980mp} including subtleties associated with this specific case for future considerations.

\ack
The author would like to thank M. Shifman for his guidance on this project.

\end{document}